\documentclass[12pt]{iopart}
\expandafter\let\csname equation*\endcsname=\relax
\expandafter\let\csname endequation*\endcsname=\relax
\usepackage{amsmath}
\usepackage{iopams}
\usepackage{comment}

\newcommand*{\mom}{\bi{m}}

\newcommand*{\abs}[1]{\lvert #1 \rvert}
\renewcommand*{\a}{\bi{a}}
\renewcommand*{\u}{\bi{u}}
\newcommand*  {\x}{\bi{x}}
\renewcommand*{\v}{\bi{v}}
\newcommand*  {\w}{\bi{w}}

\newcommand*  {\R}{\bi{R}}
\newcommand {\av}{\bi{\Omega}}
\renewcommand {\e}{\bi{e}}
\newcommand*  {\grad}{\bi{\nabla}}
\newcommand*  {\Cdot}{\bi{\cdot}}
\newcommand*  {\Eta}{\bfeta}
\renewcommand*{\Xi}{\bxi}

\newcommand{\dom}{M}

\newcommand{\eps}{\varepsilon}
\newcommand{\diff}{\mathcal{D}}

\newcommand{\vm}{\bi{m}}
\newcommand{\dlu}{\mom^{\sharp}}

\newcommand{\hdom}{M^{\prime}}
\newcommand{\T}{\mathbb{T}}
\renewcommand*{\d}{\rmd}
\renewcommand*{\P}{{\mathbb P}}

\newcommand*{\Div}{\operatorname{div}}
\newcommand{\Lie}{\operatorname{\mathcal{L}}}

\newcommand{\argmin}{\operatorname*{arg\,min}}
\newcommand{\Ds}{\mathcal{D}}

\newcommand{\vf}{\mathfrak{X}}

\newcommand{\vfd}{\vf_{\operatorname{div}}}
\newcommand{\RR}{\mathbb{R}}
\newcommand{\fluct}{{\eps, \beta}}
\newcommand{\sfluct}{{s,\beta}}

\begin{document}
\title{Geometric Lagrangian averaged Euler-Boussinesq and primitive equations}

\author{Gualtiero Badin}
\address{Center for Earth System Research and Sustainability (CEN), University of Hamburg, D-20146 Hamburg, Germany}
\ead{gualtiero.badin@uni-hamburg.de}
\author{Marcel Oliver}
\address{School of Engineering and Science, Jacobs University, D-28759 Bremen, Germany} 
\ead{oliver@member.ams.org}
\author{Sergiy Vasylkevych}
\address{Center for Earth System Research and Sustainability (CEN), University of Hamburg, D-20146 Hamburg, Germany}
\ead{sergiy.vasylkevych@uni-hamburg.de}

\vspace{10pt}
\begin{indented}
	\item[] \today
\end{indented}

\begin{abstract}
	In this article we derive the equations for a rotating stratified fluid governed by inviscid Euler-Boussinesq and primitive equations that account for the effects of the perturbations upon the mean. Our method is based on the concept of geometric generalized Lagrangian mean recently introduced by Gilbert and Vanneste, combined with generalized Taylor and horizontal isotropy of fluctuations as turbulent closure hypotheses. The models we obtain arise as Euler-Poincar\'{e} equations and inherit from their parent systems conservation laws for energy and potential vorticity. They are structurally and geometrically similar to Euler-Boussinesq-$\alpha$ and primitive equations-$\alpha$ models, however feature a different regularizing second order operator.  
\end{abstract}



	

\section{Introduction}
The goal of this article is to derive a self-consistent system of equations for the flow of an inviscid stratified fluid that accounts for the effects of the perturbations upon the mean. To this end, each realization of the flow is decomposed into the mean part and a small fluctuation. The effects of the fluctuations on the mean flow is then described by an appropriate mean flow theory. 

The definition of mean flow is subject to choice and can be approached either from Eulerian or Lagrangian point of view (e.g. \cite{Badin,Franzke}). Eulerian mean theories define the mean flow by averaging fluid velocities at each spatial location. Lagrangian averaging describes the mean and deviations in terms of particle trajectories. Since most conservation laws in fluids hold for each fluid parcel, the Lagrangian point of view is more natural whenever the conservation laws are emphasized and this is the path that we pursue in the article. 

The equations for the mean flow will be expressed in the form of $\alpha$-models for turbulence, which will be derived making use of geometric Lagrangian averaging. The original $\alpha$-models \cite{HoMaRa98,HoMaRa02} rely on the observation that replacing the kinetic energy  with its $H^1$ counterpart
\begin{eqnarray} \label{e.rega}
\int_{\dom} |\u |^2 \d \x \mapsto \int_{\dom} |\u |^2 +\alpha^2 | \grad \u |^2 \d  \x 
\end{eqnarray}
in the fluid Lagrangian results in filtering, which moderates the growth of instabilities at wave numbers larger than $O(1/\alpha^2)$, has a negligible impact on large scales and geophysical balances, and preserves the Hamiltonian structure of the equations and accompanying conservation laws.  Motivated by this observation, a number of turbulence models were proposed, including among others,  Euler-$\alpha$, Navier-Stokes-$\alpha$ (NS-$\alpha$), Euler-Boussinesq-$\alpha$ (EB-$\alpha$), and primitive equations-$\alpha$ (PE-$\alpha$) \cite{Chen,HoMaRa98,HoMaRa02,Holm02,Holm02b,MS03}. 
These models were originally derived as abstract Euler-Poincar{\'e} equations arising from the above regularization in the appropriate context. Later on Euler-$\alpha$ and NS-$\alpha$ were re-derived via Lagrangian averaging, further justifying their use for turbulence modelling.

The concept of Lagrangian averaging was introduced by Andrews and McIntyre in their seminal paper \cite{AM78} (see also e.g. \cite{AM78b,Buehler,RS06a,RS06b,SR10,Salmon13,Salmon16,SR14,W15,X15}). According to the generalized Lagrangian mean (GLM) theory of Andrews and McIntyre, the mean flow is defined by Reynolds averaging particle positions in Eucledean space for each Lagrangian label $\a$, i.e. 
\begin{eqnarray} \label{e.glm}
	\Eta(\a,t) \equiv \langle \Eta^{\Xi}(\a,t)\rangle \,,
\end{eqnarray}
where $\Eta$ is the Lagrangian mean flow, $\{\eta^{\Xi} \}$ is the ensemble of the flows to average, and $\langle \cdot \rangle$ is the ensemble average.  The mean Lagrangian velocity $\u$ is then determined from the mean flow by time differentiation
\begin{eqnarray} \label{e.uglm}
	\u (\x,t)=\frac{\partial}{\partial t} \Eta (\a,t) \qquad \mbox{with} \qquad  \a=\Eta^{-1}(\x,t) \,.
\end{eqnarray}
In order to obtain a closed system of equations for the mean, one must supplement the GLM equations with closure laws governing the evolution of fluctuations. 

A number of models for turbulent flow, including among others, Holm's Lagrangian Averaged Euler (LAE), Lagrangian mean Euler-Boussinesq (LMEB) and Lagrangian mean motion (LMM) equations, Euler-$\alpha$, and Navier-Stokes-$\alpha$ (NS-$\alpha$) EB-$\alpha$ and PE-$\alpha$ \cite{Chen,HoMaRa98,Holm99,Holm02,Holm02b,MS03}, were derived combining GLM with the generalized Taylor hypothesis, which postulates that the fluctuations vector field $\Xi \equiv \Eta^{\Xi}-\Eta$ is Lie transported by the mean flow, i.e. 
\begin{eqnarray} \label{e.taylor}
 \partial_t \Xi + \grad_{\u} \Xi - \grad_{\Xi} \u =0 \,,
\end{eqnarray}  
where we employ the geometric notation $\nabla_{\u} \Xi$ to express the derivative of a field $\Xi$ in the direction of $\u$, so that $\nabla_\u \Xi \equiv (\u \cdot \nabla) \Xi$.

The major drawback of the original GLM theory is its reliance on Euclidean space structure to define the mean flow. Not only it prevents a seamless transfer of the theory to the manifold context, but also potentially breaks the incompressibility of the mean even if all the averaged flows are incompressible.
This poses an unsatisfying dilemma: one either has to accept small compressibility of order $O(|\Xi|^2)$ or impose additional {\em ad hoc} constraints on second order fluctuations, which restore incompressibility. The first approach is taken in \cite{Holm99}, while the latter was used in \cite{MS03} to derive the Euler-$\alpha$ model. 

 Recently, Gilbert and Vanneste \cite{GV} proposed an alternative method of Lagrangian averaging that accounts for the underlying geometry. In the resulting geometric GLM (GGLM), rather than averaging particle trajectories, one averages the flows themselves, relying on the manifold structure of a diffeomorphism group in doing so. More precisely, the GGLM mean flow is defined as the Riemannian centre of mass on an appropriately chosen diffeomorphism group $\diff$, 

\begin{eqnarray}
\Eta= \argmin_{\Psi \in \diff } \langle d^2(\Psi,\Eta^{\Xi}) \rangle \,,
\end{eqnarray}
where $d$ is the geodesic distance on $\diff$. Then, the fluctuations, rather than being vector fields, are the diffeomorphisms connecting individual realizations with the mean, so that
\begin{eqnarray}
\Eta^{\Xi}=\Xi \circ \Eta \,.
\end{eqnarray}
We remark that if one chooses a volume preserving diffeorphism group $\diff_\mu$ as a configuration space, the mean flow is automatically volume preserving. 

Another important choice arising in turbulence modelling is the treatment of Taylor diffusivity tensor $\kappa= \langle \Xi \otimes \Xi \rangle$. An attractive property is isotropy, 
\begin{eqnarray} \label{e.iso}
\langle \Xi \otimes \Xi \rangle = \alpha I \,,
\end{eqnarray}
where $I$ is the identity matrix, as it significantly simplifies the equations of motion, leading, among others, to classical Euler- and NS-$\alpha$ equations. Combining GGLM with Taylor's hypothesis and isotropy yields the Euler-$\alpha$,  Camassa-Holm and EPdiff equations \cite{O17,OV18} as Euler's and Burgers' mean flow, respectively. An alternative approach is to study a coupled dynamical system for the mean velocity and Taylor diffusivity tensor, where the issue of proper initialization for $\kappa$ is a significant one. The examples of this approach are LMEB and anisotropic Euler-$\alpha$ equations \cite{Holm99,MS03}. 

The stratified analogues of the classical Euler-$\alpha$ equations, the Euler-Bousinessq-$\alpha$ (EB-$\alpha$) and primitive equations-$\alpha$ (PE-$\alpha$) were, to the best of our knowledge, never derived as a turbulence model, although they were proposed and studied numerically \cite{HoMaRa98,HoMaRa02,HolmWingate}. This is most likely due to the fact that isotropy \eqref{e.iso} is not compatible with rigid lid boundary conditions, natural for ideal geophysical flows. We overcome this issue by noting that rigid lid boundary conditions are compatible with 
 horizontal isotropy 

\begin{eqnarray}
	\kappa = \begin{pmatrix} 
	1&0&0\\
	0&1&0\\
	0&0&0 
	\end{pmatrix}
\end{eqnarray}
on a horizontal strip. Notice that in the atmosphere and in the ocean, the term ``horizontal'' indicates a flow either along geopotential or isopycnal surfaces, respectively. In GGLM, horizontal isotropy must be interpreted accordingly.   

Complimenting GGLM with the generalized Taylor hypothesis and horizontal isotropy as closure assumptions, we derive the models for the mean flow governed by the inviscid Euler-Boussinesq and primitive equations. We will call the respective models the horizontally isotropic Lagrangian averaged Euler-Boussinesq (HILAEB) and the horizontally isotropic Lagrangian averaged primitive equations (HILAPE). HILAEB and HILAPE are structurally and geometrically similar to EB-$\alpha$ and PE-$\alpha$, in particular they arise as Euler-Poincar{\'e} equations on the diffeomorphism group, and inherit conservation laws from their respective parent systems. However, we replace the full 3D Laplacian in EB-$\alpha$ and PE-$\alpha$ by a horizontal one, 
\begin{eqnarray}
\Delta_h = \partial_{x_1 x_1}+ \partial_{x_2 x_2}. 
\end{eqnarray}

The discussion whether $\alpha$-models are good turbulence models  (see, for instance, \cite{Aizinger15, HolmJeff05, MZ}), is beyond the scope of the present article. Instead, the derivation exposes a set of assumptions, sufficient to justify the use of the models. 

The article has the following structure.  In section \ref{s.primitive} we introduce inviscid Euler-Boussinesq,  primitive equations, and the notation to be used throughout the article. In section \ref{s.plan} we review the GGLM concept and describe our method in a general setting. Section \ref{s.closure} is devoted to the closure hypothesis. In section \ref{s.lagr} we recall the variational principles for EB and PE, followed by derivation of HILABE in section \ref{s.ilaeb} and HILAPE in section \ref{s.hilape}, respectively. We conclude by the discussion of results is section \ref{s.discuss}.

\section{Inviscid rotating Euler-Boussinesq and primitive equations}
\label{s.primitive}
Throughout the paper, we adapt the following notation: bold-face
letters always denote three-component objects, while regular typeface
is used for their horizontal parts; e.g., we write $\u = (u_1, u_2, u_3)^T = (u, u_3)^T$ for the full three-dimensional fluid velocity, $\x = (x_1, x_2, z)^T = (x, z)^T$, for spatial coordinate, $\a=(a_1,a_2,a_3)$ for particle labels, and
$\grad = (\partial_{x_1}, \partial_{x_2}, \partial_z) =
(\nabla, \partial_z)$ for gradients. Further, we write $u^\bot=(-u_2,u_1)^T$ to denote the counter
clockwise rotation of $u$ by $\pi/2$,  $\u_h = (u,0)^T$ the projection of $\u$ onto the horizontal coordinate plane, and $\u^\bot=\e_z \times \u= (u^\perp, 0)^T$. We will also write $\Delta_h \equiv \partial_{x_1 x_1}+\partial_{x_2 x_2}$ for the horizontal Laplacian to distinguish it from full 3-D Laplace operator $\Delta \equiv \Delta_h + \partial_{zz}$. 

With  such notation in place, the inviscid Euler-Boussinesq equations in a rotating reference frame then read
\begin{subequations}
	\label{e.EB}
	\begin{eqnarray}
	\partial_t \u +  \grad_{\u} \u + \av \times \u + \theta \e_z +\grad p=0 \, \label{e.ebmom}\\
	\label{e.ebadv}
	\partial_t \theta + \grad_{\u} \theta = 0 \,, \\
	\label{e.ebdiv}
	\grad \, \Cdot \, \u = 0 \,,
	\end{eqnarray}
\end{subequations}
where $\av (x) / 2$ is the local angular velocity vector.  
The EB equations approximate the motion of stratified inviscid incompressible Newtonian fluid by neglecting the variations in fluid density everywhere except in the gravity term. They are physically relevant for large- and mesoscale flows both in atmosphere and ocean. In the former case, $z$, $p$ and $\theta$ stand for pressure, geopotential and potential temperature, while in the context of ocean flows they represent depth, pressure and buoyancy, respectively. 

For simplicity, we consider  equations in the strip $\dom=\hdom \times [0,H]$ of constant height $H$ with no mass flux conditions  
\begin{eqnarray} \label{e.BC}
	u_3=0\, \mbox{ for } z=0 \mbox{ and } z=H		
\end{eqnarray}
 on vertical boundaries. The horizontal projection of the strip, $\hdom$, is assumed to be either a plane, $\hdom=\RR^2$, a double periodic domain $\hdom=\T^2$, or an infinite cylinder $\hdom=\RR \times \T$. In all cases we assume sufficient decay at infinity, so that volume integrals defining the Lagrangians below are finite and we can freely integrate by parts in the horizontal variables without incurring boundary terms. 
 
 The primitive equations are a further simplification of the EB system that arises by neglecting the vertical component of the Coriolis force and imposing hydrostatic balance in the vertical momentum equation. The form of the primitive equations considered here reads
 \numparts
 	\label{e.primitive}
 	\begin{eqnarray}
 	\partial_t \u_h +  \grad_{\u} \u_h + f \u^\bot + \theta \e_z +\grad p=0 \,,\\
 	\label{e.peadv}
 	\partial_t \theta + \grad_{\u} \theta = 0 \,, \\
 	\label{e.pediv}
 	\grad \,\Cdot \, \u = 0 \,,
 	\end{eqnarray}
 	\endnumparts
 where $f=f(x)$ is the Coriolis frequency. 
 
 \section{Turbulence models via Lagrangian averaging for Hamiltonian systems} \label{s.plan}
 This section describes the general procedure we use to derive turbulence models for flows governed by EB and primitive equations. The derivation itself is postponed to section \ref{s.derivation}. The procedure is applicable to arbitrary Hamiltonian systems on diffeomorphism groups.  We refer the reader to \cite{O17,OV18} for further examples of such applications.   
 
 The cornerstone of our approach is the concept of geometric generalized Lagrangian mean introduced by Gilbert and Vanneste \cite{GV}. Suppose $\{\beta\}$ is an arbitrary index set, $\{\Eta_\beta\}$ is an ensemble of flows parametrized by $\beta$ and $\langle f_\beta \rangle$ denotes the average of scalar quantity $f$ over $\{\beta\}$. The precise way in which this averaging is defined does not matter as long as it commutes with time-differentiation and spatial integration 
\begin{eqnarray}
\left\langle \int_\dom f_\beta (x) \, \d \x \right\rangle = \int_\dom \langle f_\beta (x) \rangle \, \d \x 
\end{eqnarray}
and the chosen closure assumptions are satisfied with respect to the induced notion of the Lagrangian mean flow. The GGLM exploits the fact that flow maps $\eta_\beta$ are points on an infinite dimensional Riemannian manifold, namely an appropriately chosen diffeomorphism group, and therefore their ensemble average can be defined as the geometric centre of mass on that group. 

To make this idea more precise, we choose as the configuration space $\Ds=\Ds^s_\mu (M)$ the group of volume and orientation preserving diffeomorphisms of Sobolev class $s>5/2$ which leave the boundary invariant, 
\begin{eqnarray}
\Ds^s_\mu (M) \equiv \{ \Eta \in H^s( M) \, | \, \Eta^{-1} \in H^s(M) \,, \Eta(\partial M)= \partial M \,, \det \grad \Eta =1  \} ,
\end{eqnarray}

Let $\vfd \equiv \vfd^s(\dom)$ be the space of $H^s$ divergence free vector fields on $\dom$. Then, following \cite{Palais,EM69}, $\Ds^s_\mu$ is a smooth infinite dimensional manifold with tangent bundle 
\begin{eqnarray} \label{e.tangentbundle}
	T_{\Eta} \Ds^s_\mu (\dom) = \{ \u \circ \Eta \, | \, \u \in \vfd^s(\dom) \,, u_3=0 \mbox{ on } \partial M  \} \,.
\end{eqnarray}
Equipping $T_{\Eta} \Ds$ with $L^2$ inner product 
\begin{eqnarray} \label{e.l2metric}
(\v \circ \Eta , \w \circ \Eta)= \int_{\dom} \v \cdot \w \, \d \x \quad  \mbox{ for all } \v, \w \in \vf	
\end{eqnarray}
turns $\Ds$ into a weak Riemmanian manifold. Then, one can define the geodesic distance between maps $\bpsi, \bphi \in \Ds$ via
\begin{eqnarray} \label{e.metric-full}
d^2(\bphi,\bpsi) 
= \inf\limits_{\substack{\bgamma_s:[0,1] \rightarrow \Ds \\ \bgamma_0=\bphi \,, \bgamma_1=\bpsi}} \int_0^1  (\bgamma_s^\prime,\bgamma_s^\prime)  \, \d t \,,
\end{eqnarray} 
where $\bgamma_s$ is a geodesic in $\Ds$ connecting $\bphi$ and $\bpsi$ with prime here and further on denoting derivative with respect to fictitious time $s$ parametrizing the geodesics. 

We remark that $\Ds$ is not geodesically complete with respect to the $L^2$ metric, therefore the above  definition is only a formal one. Nevertheless, in case when $M$ is compact, $\Ds$ has a well-defined exponential map \cite{EM69}, so that \eqref{e.metric-full} is defined in a neighbourhood of the identity and, by right invariance of the metric, in the neighbourhood of any $\Eta \in \Ds$.

Now we note that if $\{ \Eta_{\beta} \}$ is an ensemble of sufficiently smooth EB or PE flows with no-mass-flux boundary conditions, then, necessarily, $\{\Eta_{\beta} \} \subset \Ds$.  Following Gilbert and Vanneste, the Lagrangian  mean flow $\langle \Eta_{\beta} \rangle_L$ is defined as the ensemble's geometric centre of mass in the volumorphisms group, i.e.  
\begin{eqnarray}
\Eta=\langle \Eta_{\beta} \rangle_L \equiv \argmin_{\bPsi \in \diff } \langle d^2(\bPsi,\Eta_\beta) \rangle \,.
\end{eqnarray}

Then a fluctuation is defined as a diffeomorphism $\Xi_\beta$ connecting the mean with the corresponding realization, so that
\begin{eqnarray}
\Eta_\beta = \Xi_\beta \circ \Eta \,.
\end{eqnarray}
The Lagrangian mean velocity field $\u$ is recovered by differentiating the mean flow in time, i.e. 
\begin{eqnarray}
\dot \Eta \equiv \partial_t \Eta(\a,t) = \u \circ \Eta (\a,t) \,. 
\end{eqnarray}

This definition has three advantages.  First, it is geometrically
intrinsic.  Second, the mean flow automatically satisfies
no-mass-flux boundary conditions and is incompressible by virtue of
being a member of $\Ds$.  Third, for any set of quantities
$\{q_\beta\}$ that are materially conserved by the respective flows
$\Eta_\beta$, their Lagrangian average
\begin{eqnarray}
	\bar{q}^L = \langle \Xi^{*}_\beta q_\beta \rangle \,, 
\end{eqnarray}
where the star denotes the pull-back operation, is materially
conserved by the mean flow $\Eta$.  In particular, the mean motion
will have a well-defined potential vorticity (PV) conservation law,
provided each of the averaged flows has one. As usual, to obtain a
turbulence model, one needs to supply closure assumptions on
statistical properties and on the evolution of fluctuations that
result in a closed system of equations for the mean flow.

The drawback of the above construction is that the equations governing the Lagrangian mean velocity $\u$ is difficult to write explicitly. For Hamiltonian systems this difficulty can be overcome by combining the GGLM concept with averaging of the Lagrangian, as is done in \cite{MS03}. Consider a Hamiltonian system on $\Ds$ defined by a Lagrangian $L$, i.e., the system whose flows are stationary points of the action 
\begin{eqnarray}\label{e.action1}
 S=\int_0^t L(\Eta,\dot \Eta) \d t \,,  \qquad \Eta \in \Ds \,,
\end{eqnarray}
with respect to variation of $\Eta$ fixed at the temporal end points. In section \ref{s.lagr} we recall how the EB and primitive equations fit into this variational framework. 

Let $\eps>0$ be a fixed small parameter describing the amplitude of fluctuations, let $\Eta_\fluct$ be a single realization from an ensemble of turbulent flows of the system, and let $\Eta=\langle \Eta_\fluct \rangle_L$ be the Lagrangian mean of the ensemble in the GGLM sense. We expand $\left \langle L(\Eta_{\fluct},\dot \Eta_{\fluct}) \right \rangle $ in powers of $\eps$, expressing the result in terms of the mean flow and statistical properties of the fluctuations, and impose closure conditions such that 
\begin{eqnarray} \label{e.meanl}
\left\langle L(\Eta_\fluct, \dot \Eta_\fluct) \right \rangle=\bar L(\Eta, \dot \Eta)+O(\eps^3). 
\end{eqnarray}
The averaged Lagrangian $\bar L$ then yields averaged model equations via the variational principle. 

\section{Closure assumptions} \label{s.closure}
By construction, fluctuations $\Xi_{\fluct}$ are volumorphisms such that
\begin{eqnarray}
	\Eta_\fluct=\Xi_\fluct \circ \Eta \,.
\end{eqnarray}
Let $\Xi_{\sfluct}$, where $0 \leq s \leq \eps$ and $\Xi_{0, \beta} = \operatorname{id}$, be the geodesics in $\Ds$ connecting $\Eta$ with realizations $\Eta_{\fluct}$. Define the fluctuation vector fields $\w_\sfluct$ by 
\begin{eqnarray}
	\Xi_\sfluct^\prime= \w_\sfluct \circ \Xi_\sfluct \,.
\end{eqnarray}
Gilbert and Vanneste \cite{GV} show that $\w_\sfluct$ satisfy the Euler equations  in fictitious time $s$,
\
\begin{subequations}\label{e.transportvol}
	\begin{eqnarray} 
	\w_\sfluct^\prime + \nabla_{\w_\sfluct} \w_\sfluct \,  + \grad \phi_\sfluct =0 \,,
	\label{e.transportvol.a}
	\end{eqnarray}
	together with a constraint on the initial condition
	\begin{eqnarray}
	\langle \w_\beta \rangle = 0 \,, 
	\label{e.transportvol.b}
	\end{eqnarray}
\end{subequations}
where the absence of $s$ index means evaluation at $s=0$. The potentials $\phi_{\sfluct}$ in \eqref{e.transportvol} are uniquely (up to a constant) determined by the requirement to maintain incompressibility and boundary conditions. By the Hodge decomposition, they satisfy Poisson's equation 
\begin{subequations} \label{e.poissonphi}
	\begin{eqnarray}
\Delta \phi_{\sfluct}=- \Div (\nabla_{\w_{\sfluct}}\w_{\sfluct })	\quad \text{in } M	
	\end{eqnarray}
with matching Neumann boundary conditions
\begin{eqnarray}
\frac{\partial \phi}{\partial z} = - (\nabla_{\w_{\sfluct}}\w_{\sfluct }) \cdot \e_z \quad \text {on } \partial M. 
\end{eqnarray}
\end{subequations}

As before, let $\u$ denote the Eulerian velocity generating the Lagrangian mean flow $\Eta$. Both for the Euler-Boussinesq and the primitive equations we choose the generalized Taylor hypothesis
\begin{subequations} \label{e.closure}
	\begin{eqnarray}
	\label{e.gentaylor}
	\dot \w_\beta + \Lie_{\u} \w_\beta=0 \, 
	\end{eqnarray}
	and horizontal isotropy of fluctuations,
	\begin{eqnarray} \label{e.hisotropy}
	\langle \w_\beta \otimes \w_\beta \rangle = \begin{pmatrix}
	1&0&0 \\
	0&1&0 \\
	0&0&0 
	\end{pmatrix} \,,
	\end{eqnarray}
\end{subequations}
as closure assumptions.  

The generalized Taylor hypothesis, which implies that first order fluctuation vector fields are transported by the mean flow, is the usual assumption for derivation of Lagrangian turbulence models \cite{Holm99,Chen,Holm02,Holm02b,MS03,O17}. 
On domains without boundary, the Euler-$\alpha$ equations were derived using full isotropy of fluctuations,
	\begin{eqnarray} \label{e.isotropy}
	\langle \w_\beta \otimes \w_\beta \rangle = I \,.
	\end{eqnarray}
However, it has been noted already in \cite{MS03} that \eqref{e.isotropy} is incompatible with no-mass-flux boundary conditions, hence anisotropic Lagrangian averaged equations must be considered on domains with boundaries. While our methods allow for derivation of anisotropic equations as well, we note that the horizontal isotropy is consistent with boundary conditions considered here. We further note that a horizontal initial condition for the Euler equation \eqref{e.transportvol.a} generally results in a nearly horizontal flow map for $s\leq \eps$. Therefore a horizontal vector field $\w_\beta$ generates nearly horizontal fluctuation $\Xi_{\fluct}$. As remarked earlier, in geophysical stratified flows the vertical scale of motion is typically significantly smaller than horizontal one, so that the horizontal isotropy assumption is reasonable from physical standpoint. 

\section{Variational principles for Euler-Boussinesq and primitive equations} \label{s.lagr}
In this section we recall how EB and PE arise from a variational principle on $\Ds$. Let $\Eta$ denote the flow of a time-dependent Eulerian velocity field $\u$, i.e. 
\begin{eqnarray}\label{e.udef}
\dot \Eta \equiv \partial_t \Eta(\a,t) = \u \circ \Eta (\a,t) \,. 
\end{eqnarray}
with $\Eta(\cdot, 0)$ being an identity map. The advection equation \eqref{e.ebadv} is then equivalent to 
\begin{eqnarray}\label{e.thetadef}
\theta \circ \Eta = \theta_0 \,,
\end{eqnarray}
where $\theta_0$ is the initial potential temperature distribution. Let $\R$ be a vector potential for the angular velocity $\av$ and consider the Lagrangian  
\begin{eqnarray}
L(\Eta,\dot \Eta)=\int_{\dom} \tfrac{1}{2}| \dot \Eta |^2  + \R \circ \Eta \cdot \dot \eta - \theta_0 \eta_3 \,  \d \a \, 
\end{eqnarray}

One can  show that $\u$ and $\theta$ satisfy EB system \eqref{e.EB} if and only if the flow map $\Eta$ is a stationary point of the action,
\begin{eqnarray}\label{e.action}
\delta S =0, \qquad \, S=\int_0^t L(\Eta,\dot \Eta) \d t \,, \qquad \Eta \in \Ds \,,
\end{eqnarray} 
with respect to variations of $\Eta$ fixed at temporal end points. 

Let $\vf=T_{\operatorname{id}} \Ds$. Because of \eqref{e.tangentbundle}, we can write $\delta \Eta = \w \circ \Eta$, with $\w \in \vf$.  Using the particle relabelling symmetry,
\begin{eqnarray}\label{e.ell} 
L(\Eta,\dot \Eta)=\ell (\u,\theta)\equiv  \int_\dom \tfrac{1}{2}|\u |^2 + \R \cdot \u - \theta  z \, \d \x \,, 
\end{eqnarray}
where, as always, $\u$, $\theta$, and $\Eta$ are related by \eqref{e.udef}-\eqref{e.thetadef}, the variational principle leading to the EB equations can be restated in purely Eulerian terms as 
\begin{eqnarray}
\label{e.redaction}
\delta \int^{t}_0 \ell (\u, \theta) \, \d t= 0, 
\end{eqnarray}
subject to variations in $\u$ and $\theta$ obeying the Lin constraints 
\begin{subequations}  \label{e.Lin}
	\begin{eqnarray}
	\delta \u = \dot{\w} + [\u, \w] \,, \\
	\delta \theta+\grad_{\w} \theta = 0 \,.
	\end{eqnarray}
\end{subequations}
Lin constraints reflect the fact that even though the reduced Lagrangian $\ell$ depends only on Eulerian quantities, the variations are still taken with respect to the flow map $\Eta$. 

The Euler-Poincar\'e theorem for continua \cite{HoMaRa98,HSS09} states that Euler-Poincar\'e equations for \eqref{e.action} or, equivalently, for \eqref{e.redaction}, \eqref{e.Lin}, are given by 
\begin{eqnarray}
\label{e.ep}
\int_\dom (\partial_t + \mathcal{L}_{\u}) \mom \Cdot \w 
+ \frac{\delta \ell}{\delta \theta} \, \mathcal{L}_{\w} \theta \,  
\d \x = 0 \mbox{ for all } \w \in \vf  \,, 
\end{eqnarray}
where $\Lie$ denotes the Lie
derivative and $\vm$ is the momentum one-form $\vm  = \frac{\delta \ell}{\delta \u} \,
$. Translated to the language of vector calculus, up to a vanishing term $\int_{\dom} \grad (\dlu \cdot \u) \cdot \w \, 
\d \x $, \eqref{e.ep} reads 
\begin{eqnarray} \label{e.epvec}
\int_\dom \left (\partial_t  \dlu+ (\grad \times \dlu ) \times \u
+  \frac{\delta \ell}{\delta \theta} \grad \theta \right)\cdot \w \,  \,  
\d \x = 0 \mbox{ for all } \w \in \vf  \,, 	
\end{eqnarray}
where $\dlu$ is a vector field dual to $\mom$ with respect to $L^2$
metric. 
Equation \eqref{e.epvec} expresses that the term in parenthesis is
$L^2$-orthogonal to arbitrary sufficiently smooth divergence-free
vector fields which are tangent to the boundary.  The Hodge
decomposition theorem then implies that this term must be  a gradient.
The EB momentum equation \eqref{e.ebmom} follows by noting that   
\begin{eqnarray}
\dlu = \R +  \u \,,
\qquad \frac{\delta \ell}{\delta \theta} = -z \,,
\end{eqnarray}
and using the vector identity 
\begin{eqnarray}
(\grad \times \u ) \times \u = \grad_{\u} \u - \tfrac{1}{2} \grad | \u |^2 \,.
\end{eqnarray}

Analogous to EB system, the primitive equations arise from variational principle \eqref{e.action} on the diffeomorphism group $\Ds$ with Lagrangian 
\begin{eqnarray}\label{e.pelagr}
L^{P} (\Eta, \dot \Eta)= \int_{\dom} \tfrac{1}{2}| \dot \eta |^2  + \R \circ \Eta \cdot \dot \Eta - \theta_0 \eta_3 \,  \d \a \,,
\end{eqnarray}
where $\R=(R(x),0)^T$ is the vector potential for the Coriolis parameter, i.e. $\grad \times \R=\av=f \e_z$. 

\section{Averaged EB and PE Lagrangians} \label{s.derivation}
\subsection{Euler-Boussinesq}
Now we are ready to implement the program outlined in section \ref{s.plan} using the closure conditions \eqref{e.closure}. Using notation of section \ref{s.closure}, for $0\leq s \leq \eps$ define 
\begin{eqnarray}
\Eta_{\sfluct}=\Xi_{\sfluct} \circ \Eta
\end{eqnarray} 
and let $\u_\sfluct$ and $\u$ denote Eulerian velocity fields generating $\Eta_\sfluct$ and $\Eta$, respectively, while $\theta_\sfluct$ and $\theta$ stand for corresponding potential temperature fields. We have
\begin{subequations} \label{e.uthetadef}
	\begin{eqnarray}
	\dot \Eta_\sfluct = \u_\sfluct \circ \Eta_\sfluct \,, \qquad & \dot \Eta  = \u \circ \Eta \,, \\
	\theta_\sfluct \circ{\Eta_\sfluct}=\theta_0 \,, \qquad &\theta \circ \Eta = \theta_0. 
	\end{eqnarray}
\end{subequations} 
Expand the velocity and potential temperature fields in powers of $s$, 
\begin{subequations} \label{e.u-expansion}
	\begin{eqnarray}
	\u_\sfluct
	= \u + s \, \u_\beta^\prime
	+ \tfrac12 \, s^2 \, \u_\beta^{\prime \prime} + O(\eps^3) \,, \\
	\theta_\sfluct
	= \theta + s \, \theta_\beta^\prime
	+ \tfrac12 \, s^2 \, \theta_\beta^{\prime \prime} + O(\eps^3) \,, 
	\end{eqnarray}
where we are using a uniform bound on the Taylor's series remainder, 
\end{subequations}
and substitute into EB Lagrangian to obtain
\begin{eqnarray}\label{e.l-expansion}
L_\eps &\equiv\langle L(\Eta_\fluct, \dot \Eta_\fluct)\rangle \notag \\
&=\frac12 \,
\biggl\langle
\int_\dom
\lvert \u \rvert^2
+ 2 \, \varepsilon \, \u \cdot \u_\beta^\prime
+ \varepsilon^2 \,
\bigl(\lvert \u_\beta^\prime \rvert^2 + \u \cdot \u_\beta^{\prime \prime} \bigr) \, \d \x
\biggr\rangle \notag \\
&+\biggl\langle
\int_\dom
\R \cdot (\u+\eps \u_\beta^\prime + \tfrac{\eps^2}{2} \u_\beta^{\prime \prime})-
z(\theta+\eps \theta_\beta^\prime + \tfrac{\eps^2}{2} \theta_\beta^{\prime \prime})
\, \d \x 
\biggr\rangle
+ O(\eps^3).
\end{eqnarray}

In order to obtain the mean flow Lagrangian in the closed form \eqref{e.meanl}, one needs to eliminate all quantities with primes from \eqref{e.l-expansion}. The expansion fields $\u_\beta^\prime$, $\theta_\beta^\prime$, etc. can be expressed in terms of mean quantities and fluctuations by differentiating the curves  at $s=0$. Indeed, 
the equality of partial derivatives
\begin{eqnarray}
(\dot \Eta_\sfluct)^\prime=\partial_t{(\Eta_\sfluct^\prime)}
\end{eqnarray}
yields 
\begin{eqnarray}
(\u_\sfluct^\prime+\nabla_{\w_\sfluct} \u_\sfluct)\circ \Eta_\sfluct=(\dot \w_\sfluct + \nabla_{\u_\sfluct} \w_\sfluct) \circ \Eta_\sfluct \,,
\end{eqnarray}
therefore, since $\u_{0,\beta}=\u$,
\begin{eqnarray}\label{e.up}
\u^\prime_\beta= \dot \w_\beta+\Lie_{\u} \w_\beta. 
\end{eqnarray}
Similarly, differentiating $\eqref{e.thetadef}$ yields 
\begin{eqnarray}\label{e.thetap}
\theta_\beta^\prime=-\grad_{\w_\beta} \theta \,.
\end{eqnarray}

The Taylor hypothesis \eqref{e.gentaylor} implies  
\begin{eqnarray}\label{e.up0}
\u_\beta^\prime = 0
\end{eqnarray}
and
\begin{eqnarray}
\u_\beta^{\prime \prime} &= \dot \w_\beta^\prime + \Lie_{\u} \w_\beta^\prime + \Lie_{\u^\prime} \w_\beta \notag \\
&= \dot \w_\beta^\prime + \Lie_{\u} \w_\beta^\prime \,.
\end{eqnarray}
Differentiating \eqref{e.transportvol.a} in time and replacing $\dot \w_\beta$ by $-\Lie_{u} \w_\beta$ from \eqref{e.gentaylor}, we can express $\dot \w_\beta^\prime$ in terms of $\w_\beta, \u$, and $\phi$ to obtain 
\begin{eqnarray} \label{e.upp}
\u_\beta^{\prime \prime} 
&= \grad_{\w_\beta} (\Lie_{\u} \w_\beta) \,  + \grad_{\Lie_{\u} \w_\beta}  \w_{\beta} \, 
- \Lie_{\u} 
(\grad_{\w_\beta} \w_\beta) - 
		\Lie_{\u} \grad \phi - \grad \dot \phi  \notag \\
&=\grad_{\grad_{\w_{\beta}} \w_\beta} \u - \grad_{\w_\beta} \grad_{\w_\beta} \u-\Lie_{\u} \grad \phi - \grad \dot \phi \,, 
\end{eqnarray} 
where, in order to pass to the second line in \eqref{e.upp}, we used the standard vector identities   
\begin{subequations}
	\begin{eqnarray}
	\Lie_{\u} \v=\grad_u \v - \grad_{\v} \u \\
		\grad_{\w} \grad_{\u} \v - \grad_{\u}\grad_{\w} \v + \grad_{\Lie_{\u}\w } \v = 0 
	\end{eqnarray}
for arbitrary vector fields $\u$, $\v$, and $\w$. We remark that the last identity holds since $\dom$ has no curvature. 
\end{subequations}
Finally, differentiating \eqref{e.thetap} and substituting $\w_\beta^\prime$  and $\theta_\beta^\prime$ from \eqref{e.transportvol.a} and \eqref{e.thetap}, respectively, into the resulting expression yields 
\begin{eqnarray}\label{e.thetapp}
	\theta_\beta^{\prime \prime}=\grad_{\grad_{\w_\beta} \w_\beta} \theta+\grad_{\grad \phi_\beta} \theta+\grad_{\w_\beta} \grad_{\w_\beta} \theta
\end{eqnarray}
Now, using \eqref{e.up0}, \eqref{e.thetap}, \eqref{e.upp}, and \eqref{e.thetapp}, we can eliminate the higher order derivatives from \eqref{e.l-expansion} and express the azimuthally averaged Lagrangian $L_\eps$ as 
\begin{eqnarray}\label{e.l-exp1}
	L_\eps&=\ell(\u,\theta)+\eps \left \langle \int_{\dom} z \grad_{\w_\beta} \theta \, \d \x \right\rangle \notag  \\ 
	&+ \tfrac{\eps^2}{2} \left\langle \int_\dom (\u+\R) \cdot (\grad_{\grad_{\w_{\beta}} \w_\beta} \u - \grad_{\w_\beta} \grad_{\w_\beta} \u-\Lie_{\u} \grad \phi_\beta - \grad \dot \phi_\beta )  \, \d \x \right\rangle \notag \\ 
	&-\tfrac{\eps^2}{2}\left\langle \int_\dom z (\grad_{\grad_{\w_\beta} \w_\beta} \theta+\grad_{\grad \phi_\beta} \theta+\grad_{\w_\beta} \grad_{\w_\beta} \theta )  \, \d \x \right\rangle +O(\eps^3) \notag \\
	&=\ell(\u,\theta) + \tfrac{\eps^2}{2} \left\langle \int_\dom (\u+\R) \cdot (\grad_{\grad_{\w_{\beta}} \w_\beta} \u  - \grad_{\w_\beta} \grad_{\w_\beta} \u-\Lie_{\u} \grad \phi_\beta )  \, \d \x \right\rangle \notag \\
	&-\tfrac{\eps^2}{2}\left\langle \int_\dom z (\grad_{\grad_{\w_\beta} \w_\beta} \theta+\grad_{\grad \phi_\beta} \theta+\grad_{\w_\beta} \grad_{\w_\beta} \theta)  \, \d \x \right\rangle +O(\eps^3) \,,
\end{eqnarray}
where $\int_\dom (\u+\R) \cdot \nabla \dot \phi_\beta \, \d \x =0 $ since divergence free fields are $L^2$-orthogonal to gradients while order $\eps$ term vanishes due to $ \langle \w_\beta \rangle=0$.

We further simplify \eqref{e.l-exp1} by integrating by parts keeping in mind that $\u$ and $\w_\beta$ are divergence free, and using the horizontal isotropy of fluctuations. Let $\v$ denote an arbitrary (not necessarily divergence free) sufficiently smooth vector field. Then 
\begin{eqnarray} \label{e.int1}
\left \langle \int_{\dom} \grad_{\grad_{\w_\beta} {\w_\beta} } \u \cdot \v \, \d \x \right \rangle &= -\left \langle \int_{\dom} w_{\beta{k}} w_{\beta{i}} \partial_{x_i} \left(\frac{\partial u_j}{\partial x_k} v_j  \right)  \, \d \x \right \rangle \notag \\
&= \int_{\dom} \partial_{x_\alpha} \left( \frac{\partial u_j}{\partial x_\alpha} v_j  \right)  \, \d \x  \notag \\
&= 0 \,,
\end{eqnarray}
where we employ Einstein's convention on summing over repeated
indices, and $\alpha$ denotes the index that runs over the horizontal
components as opposed to Roman indices running through all three
spatial coordinates.  We also
note that the horizontal isotropy of fluctuations implies the useful identity 
\begin{eqnarray} \label{e.divh}
\left \langle \int_{\dom} \w_\beta \cdot \nabla_{\w_\beta} \v \, \d \x \right \rangle
&= 
 \int_{\dom} \left \langle \w_{\beta i}  {\w_{\beta k}} \right \rangle \left( \frac{\partial v_i}{\partial x_k}\right) \, \d \x \notag \\ 
 &=\int_{\dom} \frac{\partial v_\alpha}{\partial x_\alpha} \, \d \x=\int_{\dom}  \Div \v_h  \, \d \x \notag\\
 &=0,    
\end{eqnarray}
where the last equality follows from divergence theorem since $\v_h$ is tangential to the boundary of $\dom$.

Define $ \Delta^{-1} \Div \v$  as the solution of Poisson's equation 
\begin{subequations}
	\begin{eqnarray}
	\Delta \psi  = \Div \v \quad \mbox{ in } \dom \,,
	\end{eqnarray}
with matching Neumann's boundary conditions 
	\begin{eqnarray}
	\frac{\partial \psi}{\partial z }= v_3\,  \mbox{ on }	\partial \dom \,.
	\end{eqnarray}
\end{subequations}
Then, from \eqref{e.poissonphi}, 
\begin{eqnarray}
\phi_\beta= - \grad \Delta^{-1} \Div (\grad_{\w_\beta} \w_\beta).  
\end{eqnarray}

Since the operator $\grad \Delta^{-1} \Div$ is $L^2$-symmetric,   
\begin{eqnarray} \label{e.intphi}
\left \langle \int_{\dom} \grad \phi_\beta \cdot \v \, \d \x \right \rangle&= -\left \langle\int_{\dom} \grad_{\w_\beta} \w_\beta \cdot \grad \Delta^{-1} \Div \v \, \d \x \right \rangle \notag \\ 
&=\int_{\dom} \left \langle \w_\beta \cdot \grad_{\w_\beta} \grad \Delta^{-1} \Div \v \right \rangle \, \d \x  \notag \\ 
&= 0 \,, 
\end{eqnarray} 
where we used \eqref{e.divh} in order to pass to the last line. Therefore, again integrating by parts and using \eqref{e.intphi},
\begin{eqnarray}\label{e.intphi1}
\left \langle \int_\dom \Lie_{\u} \phi_\beta \cdot \v \, \d \x \right \rangle
&=\left \langle \int_{\dom} \left( \grad_{\u} \grad \phi_{\beta}  - \grad_{\grad \phi_\beta} \u \right) \cdot \v \, \d \x \right \rangle \notag \\
&=-\left \langle \int_{\dom} \grad \phi_{\beta}   \cdot  \left( \grad_{\u}  \v + (\grad \u)^T \v \right) \, \d \x \right \rangle =0 \,.
\end{eqnarray}
Similarly, 
\begin{eqnarray} \label{e.int2}
\left \langle \int_{\dom} \grad_{\w_\beta} \grad_{\w_\beta} \u \cdot \v \, \d \x \right \rangle=- 
\int_{\dom} \nabla  \u \cdot \nabla \v \, \d \x 
\end{eqnarray}
Combining \eqref{e.int1}, \eqref{e.intphi1}, and \eqref{e.int2} yields 
\begin{eqnarray}\label{e.intv}
 \left\langle \int_\dom \v \cdot (\nabla_{\nabla_{\w_{\beta}} \w_\beta} \u  - \nabla_{\w_\beta} \nabla_{\w_\beta} \u-\Lie_{\u} \nabla \phi_\beta )  \, \d \x \right\rangle 
= \int_\dom \nabla \u \cdot \nabla \v  \, \d \x \,.
\end{eqnarray}

The potential temperature contribution to the $O(\eps^2)$ Lagrangian is 
\begin{eqnarray} \label{e.inttheta}
\left\langle \int_\dom z (\grad_{\grad_{\w_\beta} \w_\beta} \theta+\grad_{\grad \phi_\beta} \theta+\grad_{\w_\beta} \grad_{\w_\beta} \theta)  \, \d \x \right\rangle \notag \\
= - \left\langle\int_{\dom} \w_\beta \cdot \grad_{\w_\beta} (z\grad \theta) - \grad \phi_\beta \cdot (z \grad \theta) +\Div(z \w_\beta) \nabla_{\w_\beta} \theta   \, \d \x \right\rangle \notag \\
=- \int_\dom  0-0+ \left\langle w_{\beta 3} \, \w_\beta \cdot \grad \theta \right\rangle \, \d \x =- \int_\dom \left\langle w_{\beta 3} \, w_{\beta i} \right\rangle  \frac{\partial \theta}{\partial x_i}  \, \d \x =0\,,
\end{eqnarray}
where the third line follows from \eqref{e.divh}, \eqref{e.intphi}, and 
$ \langle w_{\beta 3} \, w_{\beta i} \rangle \equiv 0$. 


Inserting \eqref{e.intv} with $\v= \u +\R$ and \eqref{e.inttheta} into \eqref{e.l-exp1}, then discarding $O(\eps^3)$ terms, we find that the averaged EB Lagrangian  in reduced form reads
\begin{eqnarray}\label{e.laeblagr}
\bar \ell (\u,\theta)&= \int_{\dom} \frac1{2}| \u |^2 + \R \cdot \u -z \theta +\tfrac{\eps^2}{2} \left(  | \nabla \u |^2 + \nabla \R \cdot \nabla \u \right) \, \d \x \notag \\
 &=\int_{\dom} \frac1{2}| \u |^2 + \tilde \R \cdot \u -z \theta +\tfrac{\eps^2}{2} | \nabla \u |^2  \, \d \x \,,
 \end{eqnarray}
 where $\tilde \R$ is the effective Coriolis parameter
 \begin{eqnarray}\label{e.tildeR}
 \tilde \R = \R - \tfrac{\eps^2}{2} \Delta_h \R. 
\end{eqnarray}
Equivalently, the averaged EB Lagrangian reads
\begin{align}
\bar{L} (\Eta, \dot \Eta)= \int_{\dom} \frac1{2}| \dot \Eta |^2 +(\tilde \R \circ \Eta) \cdot \dot \Eta - \theta_0 \eta_3 \, \d \a  + \tfrac{\eps^2}{2}  \int_{\dom}  |\nabla (\dot \Eta \circ \Eta^{-1} )|^2 \circ \Eta   \, \d \a \,.
\end{align}

%

\subsection{Primitive equations}

Now we derive the averaged Lagrangian $\bar{L}^P$ for the primitive equations. The computation proceeds analogously to the Euler-Boussinesq case with EB Lagrangian $L$ replaced by PE Lagrangian \eqref{e.pelagr}. We repeat the argument in \eqref{e.l-expansion}-\eqref{e.l-exp1}, noting that $\R=\R_h$ and  $u \cdot v = \u_h \cdot \v_h=\u_h \cdot \v$ for arbitrary vector fields $\u$ and $\v$ so that
\begin{eqnarray}\label{e.lpe-exp1}
L^P_\eps&=\ell^P(u,\theta) + \tfrac{\eps^2}{2} \left\langle \int_\dom (\u_h+\R) \cdot (\grad_{\grad_{\w_{\beta}} \w_\beta} \u  - \grad_{\w_\beta} \grad_{\w_\beta} \u-\Lie_{\u} \grad \phi_\beta )  \, \d \x \right\rangle \notag \\
&-\tfrac{\eps^2}{2}\left\langle \int_\dom z (\grad_{\grad_{\w_\beta} \w_\beta} \theta+\grad_{\grad \phi_\beta} \theta+\grad_{\w_\beta} \grad_{\w_\beta} \theta )  \, \d \x \right\rangle +O(\eps^3) \notag \\
&=\ell^P(u, \theta) + \tfrac{\eps^2}{2} \int_{\dom} \nabla \u \cdot (\nabla \u_h + \nabla \R)  \, \d \x+O(\eps^3) \notag \\ 
&=\ell^P(u, \theta) + \tfrac{\eps^2}{2} \int_{\dom} |\nabla \u_h|^2 - \u \cdot \Delta_h \R \, \d \x+ O(\eps^3)\,,
\end{eqnarray}
where the second equality is due to \eqref{e.intv} and \eqref{e.inttheta}. 

Thus, the averaged PE Lagrangian is
\begin{eqnarray}\label{e.lapelagr}
\bar \ell^P (\u,\theta)= \int_{\dom} \frac1{2}| u |^2 + \tilde \R \cdot \u -z \theta  \, \d \x+\tfrac{\eps^2}{2} \int_{\dom} |\nabla \u_h |^2 \, \d \x \,,
\end{eqnarray}
or, expressed in terms of the flow map, 
\begin{align}
\bar{L}^P (\Eta, \dot \Eta)&= \int_{\dom} \frac1{2}| \dot \eta |^2 +(\tilde \R \circ \Eta) \cdot \dot \Eta - \theta_0 \eta_3 +\tfrac{\eps^2}{2} \int_{\dom}| \nabla (\dot \eta \circ \Eta^{-1}) |^2 \circ \Eta \, \d \a \,.
\end{align}

\section{Horizontally isotropic Lagrangian averaged Euler-Boussinesq equations} \label{s.ilaeb}
The horizontally isotropic Lagrangian averaged Euler-Boussinesq equations are Euler-Poincar\'e equations for the averaged Lagrangian $\bar{L}$. To shorten notation, we introduce the effective angular velocity,
\begin{eqnarray}
\tilde{\av}=\av- \tfrac{\eps^2}{2}\Delta \av \,,
\end{eqnarray}
and, borrowing the terminology from Holm \cite{Holm99, Holm02}, the circulation velocity
\begin{eqnarray}
\v = \u- \eps^2 \Delta_h \u \,.
\end{eqnarray}

Taking variations of \eqref{e.laeblagr} with Lin constraints \eqref{e.Lin}, integrating by parts and discarding full time derivatives which do not contribute to the action, we find  
\begin{eqnarray}\label{e.freeslipdl}
\delta \bar \ell &=\int_{\dom} (\u  +  \tilde \R) \cdot \delta \u -z \delta \theta  + \eps^2 \nabla \u \cdot \nabla \delta \u   \, \d \x \, \notag \\
&=  \int_{\dom}  (\v +\tilde \R) \cdot (\dot \w+ [\u,\w])  +z\grad \theta \cdot \w   \, \d \x \, \notag \\
&=- \int_{\dom}  (\dot \v + \grad_{\u} \v  +(\grad \u)^T \v + \tilde \av \times \u +\theta \e_z ) \cdot \w   \, \d \x \,
\end{eqnarray}
for arbitrary $\w \in T_{\operatorname{id}} \Ds$. Thus, from the Hodge decomposition we obtain the HILAEB momentum equation  
\begin{eqnarray}\label{e.ilaeb}
\partial_t  \v + \grad_{\u} \v + (\grad \u)^T \v +\tilde{\av} \times \u  + \theta \e_z   +\grad p=0 \,.
\end{eqnarray}

The same result could be derived by inserting  the variational derivatives 
\begin{eqnarray}
\dlu = \left(\frac{\delta \bar \ell}{\delta \u}\right)^{\sharp}=\bar \R + \v 
\qquad \mbox{and} \qquad 
\frac{\delta \ell}{\delta \theta} = -z \,,
\end{eqnarray}
into the abstract Euler-Poincar\'{e} equation \eqref{e.epvec}, which yields
\begin{eqnarray}\label{e.ilaeb2}
 \partial_t  \v + (\grad \times (\v+\tilde{\R}) ) \times \u
-  z \grad \theta  +\grad p=0 \,.
\end{eqnarray} 
After redefining  the pressure $p \rightarrow p-z \theta - \u \cdot \v $, \eqref{e.ilaeb2} simplifies to \eqref{e.ilaeb} due to the general vector identity
\begin{eqnarray}\label{e.curlxvxu}
(\grad \times \v)\times \u = \grad_{\u} \v - (\grad \v )^T \u = \grad_{\u} \v  + (\grad \u)^T v - \grad (\u \cdot \v) \,.
\end{eqnarray}

The full set of HILAEB model equations consists of the momentum equation \eqref{e.ilaeb}, the advection of potential temperature \eqref{e.ebadv}, and the incompressibility constraint \eqref{e.ebdiv} combined with no-mass-flux boundary conditions \eqref{e.BC}. 

As a Hamiltonian model, HILAEB conserve the energy 
\begin{eqnarray}
E(\u, \theta) 
&= \int_{\dom} \mom \cdot \u  \, \d \x 
- \bar \ell(\u, \theta) \notag \\ 
& = \int_{\dom} \,  \,  \tfrac{1}{2} \abs{ \u}^2 +\eps^2 \abs{\nabla \u}^2 
+\theta \, z \, \d \x \,.
\label{e.hbm}
\end{eqnarray}
The invariance of the Lagrangian $\bar L$ under particle relabelling implies the existence of materially conserved PV, 
\begin{eqnarray}\label{e.pv}
q=(\grad \times \dlu) \cdot \grad \theta
=(\tilde{\av} \times \u + \grad \times \v) \cdot \grad \theta \,.
\end{eqnarray}
The simplest way to derive the expression for PV \eqref{e.pv} is to take the wedge product of the abstract EP equation \eqref{e.ep} on the space of one-forms with $\d\theta$. We refer the reader to \cite{OV16} for the details. Alternatively, one obtains the same result after a rather tedious computation by taking the inner product of $\grad \theta$ with  the curl of \eqref{e.ilaeb}.    

We remark that PV form \eqref{e.pv} is identical to the original Euler-Boussinesq system, with $\v$ replacing fluid velocity $\u$ under the curl operator. This justifies the use of ``circulation velocity'' term for $\v$, as it is the quantity entering Kelvin's circulation theorem corresponding to material conservation of $q$, namely,
\begin{eqnarray}
\frac{\d}{\d t} \oint_{\gamma(t)} (\v+\tilde{\R}) \cdot \, \d \x = - \oint_{\gamma(t)} \theta \, \d \x \,,
\end{eqnarray}
whenever the curve $\gamma(t)$ moves with the fluid.

\section{Horizontally isotropic Lagrangian averaged primitive equations} \label{s.hilape}
To obtain  horizontally isotropic Lagrangian averaged primitive equations (HILAPE), we proceed as in \eqref{e.freeslipdl}. Computing the variation of the averaged PE  Lagrangian \eqref{e.lapelagr} yields 
\begin{eqnarray}
\delta \bar \ell^P &=  \int_{\dom}  (\v +\tilde \R) \cdot (\dot \w+ [\u,\w])  +z\grad \theta \cdot \w   \, \d \x \,  \notag \\
&=- \int_{\dom}  (\dot \v + \grad_{\u} \v  +(\grad \u_h)^T \v + \tilde f \u^{\bot} +\theta \e_z ) \cdot \w   \, \d \x \,
\end{eqnarray}
for arbitrary vector field $\w \in \vfd^s(\dom)$, where 
\begin{eqnarray}
\v=\u_h-\eps^2 \Delta_h \u_h \qquad \mbox{and} \qquad \tilde{f}=f - \tfrac{\eps^2}{2} \Delta_h f
\end{eqnarray}
are the circulation velocity and the effective Coriolis parameter, respectively. Therefore, the HILAPE momentum equation is given by  
\begin{eqnarray}\label{e.ilape}
\partial_t  \v + \grad_{\u} \v + (\grad \u_h)^T \v +\tilde f  \u^\bot  + \theta \e_z   +\grad p=0 \,.
\end{eqnarray}
Naturally, computing the Euler-Poincar\'e equations via \eqref{e.ep} with 
\begin{eqnarray}
 \vm^{\sharp}=	\left(\frac{\delta \ell}{\delta \u}\right)^{\sharp}= \u_h -\eps^2 \Delta \u_h + \tilde \R \, \quad \mbox{ and } 	\frac{\delta \ell}{\delta \theta}= -z
\end{eqnarray}
yields the same result.

Similar to the Euler-Boussinesq case, the full set of HILAPE consists of the momentum equation \eqref{e.ilape} with no-mass-flux boundary conditions \eqref{e.BC}, the advection of potential temperature \eqref{e.peadv}, and the incompressibility constraint \eqref{e.pediv}.

As an Euler-Poincar\'e system, HILAPE conserve the energy 
\begin{eqnarray} 
E(\u, \theta) 
= \int_{\dom} \,  \tfrac{1}{2} \, \bigl( \abs{ u}^2 +\eps^2 \abs{\nabla u}^2 \bigr) 
+\theta \, z \, \d x  
\label{e.ilapee}
\end{eqnarray}
and the potential vorticity $q$ on fluid particles, 
\begin{eqnarray}\label{e.ilapepv}
q=(\grad \times \dlu) \cdot \grad \theta
=(\tilde f + \grad \times \v) \cdot \grad \theta \,,
\end{eqnarray}
with corresponding Kelvin's circulation theorem
\begin{eqnarray}
\frac{d}{dt} \oint_{\gamma(t)} (\v+\tilde{\R}) \cdot \, \d \x = - \oint_{\gamma(t)} \theta \, \d \x
\end{eqnarray}
for an arbitrary closed curve $\gamma(t)$ moving with fluid velocity.

\section{Discussion}\label{s.discuss}

Inspired by the work by Gilbert and Vanneste \cite{GV} and Oliver \cite{O17}, we have derived the analogues of EB-$\alpha$ and PE-$\alpha$ via geometric Lagrangian averaging.  

In order to derive the HILAEB and HILAPE models, we have replaced the 3D isotropy with horizontal one \eqref{e.hisotropy} which is consistent with no-mass-flux on vertical boundaries. An alternative approach leads to anisotropic equations, similar to Lagrangian mean Euler-Boussinesq \cite{Holm99} and anisotropic Lagrangian averaged Euler \cite{MS03}, which introduce additional dynamics for the mean covariance tensor $\langle \w \otimes \w \rangle$.  HILAEB and HILAPE differ from EB-$\alpha$
\begin{subequations}
\begin{eqnarray}
\partial_t  \v + \grad_{\u} \v + (\grad \u)^T \v +{\av} \times \u  + \theta \e_z   +\grad p=0 \,, \\
\v = \u - \eps^2 \Delta \u \,,
\end{eqnarray}
\end{subequations}
 and PE-$\alpha$ 
 \begin{subequations}\label{e.pea}
 	\begin{eqnarray}
\partial_t  \v + \grad_{\u} \v + (\grad \u_h)^T \v + f  \u^\bot  + \theta \e_z   +\grad p=0 \,. \\
\v = \u_h -\eps^2 \Delta \u_h \,,
 \end{eqnarray}
 \end{subequations}
 respectively, 
 in the Coriolis terms and in the Laplacian operator defining the circulation velocity $\v$. 
 
 The full 3D Laplace operator appearing in EB-$\alpha$ and PE-$\alpha$ requires additional boundary conditions which do not arise naturally. Usually either no-slip or Navier-slip conditions are imposed, see\cite{Lopes, Shkoller00}.  HILAEB and HILAPE replace the Helmoltz operator defining the circulation velocity in EB-$\alpha$ and PE-$\alpha$, respectively,  by a horizontal one, which can be inverted on $M$ without imposing additional {\em ad hoc} boundary conditions. In contrast to the EB-$\alpha$ and PE-$\alpha$ equations, the filtering provided by horizontal Laplacian has no effect on vertical waves. 
 
It should be noted that while the difference from HILAEB and HILAPE from EB-$\alpha$ in the Coriolis term is insignificant and vanishes on an $f$- or $\beta$-plane, the difference in the Laplace operator defining the circulation velocity $\v$ needs to be examined in more detail.  
\section*{Acknoledgements}
We thank anonymous reviewers for helpful suggestions and encouragement. This work is a contribution to the Collaborative Research Centre TRR 181
``Energy Transfer in Atmosphere and Ocean'' funded by
the German Research Foundation.
\section*{References}

\bibliographystyle{siam}

\begin{thebibliography}{99}


	 \bibitem{Aizinger15}Aizinger V, Korn P, Giorgetta M and Reich S 2015
	 {Large-scale turbulence modelling via $\alpha$-regularisation for atmospheric simulations}, \textit{J. Turbul.} \textbf{16} 367--391
	
	\bibitem{AM78} Andrews D G and McIntyre M E 1978 An exact theory of nonlinear waves on a Lagrangian-mean flow  \textit{J. Fluid Mech.} \textbf{89} 609--646
	
	\bibitem{AM78b} Andrews D G and McIntyre M E 1978 {On wave-action and its relatives}  \textit{J. Fluid Mech.} \textbf{89} 647--664 
		
	\bibitem{Arnold}
	Arnold V I 1966 {Sur  la  g\'{e}ometrie  diff\'{e}rentielle  des  groupes  de  Lie  de  dimension  infinie  et  ses	applications \`{a} l'hydrodynamique des fluides parfaits} \textit{ Ann. Inst. Fourier} \textbf{16} 319--361

	\bibitem{Badin} 
	Badin G and Crisciani F 2018 \textit{Variational Formulation of Fluid and Geophysical Fluid Dynamics: Mechanics, Symmetries and Conservation Laws} (Springer) 
	
	\bibitem{Buehler} 
	B{\"u}hler O 2014 \textit{Waves and Mean Flows} (Cambridge: Cambridge University Press)
	
	\bibitem{Chen} 
	Chen S, Foias C, Holm D D, Olson E, Titi E S and Wynne S 1999 {A connection between the Camassa-Holm equations and turbulent flows
	in channels and pipes} \textit{Phys. Fluids} \textbf{11} 2343--2353 
	
	\bibitem{EM69} Ebin D G and Marsden J E 1970 {Groups of diffeomorphisms and the motion of an incompressible fluid} \textit{Ann. Math.} \textbf{92} 102--163
	
	\bibitem{Franzke} Franzke C L E, Oliver M, Rademacher J D M and Badin G 2018
	\textit{Multi-scale methods for geophysical flows} (submitted for publication)
	
	\bibitem{GV} Gilbert A D and Vanneste J 2018 {Geometric generalised Lagrangian-mean theories}  \textit{J. Fluid Mech.} \textbf{839} 95--134 
			
	\bibitem{Holm99} 
	Holm D D 1999 {Fluctuation effects on 3D Lagrangian mean and Eulerian mean fluid motion} \textit{Physica D} \textbf{133} 215--269
	
	\bibitem{Holm02} Holm D D 2002 {Lagrangian averages, averaged Lagrangians, and the mean effects of fluctuations in fluid dynamics} \textit{Chaos} \textbf{12} 518--530

\bibitem{Holm02b} 
Holm D D 2002 {Averaged Lagrangians and the mean effects of fluctuations
	in ideal fluid dynamics} \textit{Physica D} \textbf{ 170} 253--286

\bibitem{HolmJeff05} Holm D D, Jeffery C, Kurien S, Livescu D, Taylor M A and Wingate B A 2005 {The LANS-$\alpha$ model for computing turbulence: Origins, results, and open problems}, \textit{Los Alamos Science}  \textbf{29} 152--172		
	
	
	
	\bibitem{HoMaRa98} Holm D D, Marsden J E and Ratiu T S 1998 
	{The  Euler-Poincar\`{e}   Equations  and  Semidirect  Products
	with Applications to Continuum Theories}, \textit{Adv. Math.} \textbf{137} 1--81. 
 
	
	\bibitem{HoMaRa02} Holm D D, Marsden J E and Ratiu T S 2002 {The Euler-Poincar\`{e} equations in geophysical fluid  dynamics}, \textit{Large-Scale Atmosphere-Ocean Dynamics} vol~1 (Cambridge: Cambridge University Press) pp 251--300 

\bibitem{HSS09}
Holm D D, Schmah T and Stoica C 2009 \textit{Geometric Mechanics and Symmetry: from Finite to Infinite dimensions} (Oxford University Press) 	

\bibitem{HolmWingate} Petersen M R, Hecht M W, Holm D D and Wingate B A 2008 {Implementation of the LANS-$\alpha$ turbulence model in a primitive equation ocean model}, \textit{J. Comp. Phys.} \textbf{227} 5691--5716	
	
	\bibitem{Lopes}
	Lopes Filho M C, Nussenzveig-Lopes H J, Titi E S and Zang A, 2015 {Convergence of the 2D Euler-$\alpha$ to Euler equations in the Dirichlet case: Indifference to boundary layers}, \textit{Physica D} \textbf{292--293} 51--61 
	
\bibitem{MZ} Porta Mana P and Zanna L 2014 Toward a stochastic parameterization of ocean mesoscale eddies \textit{Ocean Model.} \textbf{79} 1--20 	
	
	\bibitem{MS03} Marsden J E and Shkoller S 2003 {The anisotropic Lagrangian averaged Euler and Navier-Stokes equations} \textit{Arch. Rat. Mech. and Analysis} \textbf{166} 27--46
	
	\bibitem{O17} Oliver M 2017 {Lagrangian averaging with
			geodesic mean} \textit{Proc. Royal Soc.} \textbf{473}  doi:10.1098/rspa.2017.0558.
	
	\bibitem{OV16} Oliver M and Vasylkevych S 2016 {Generalized large-scale semigeostrophic approximations for the $f$-plane primitive equations} \textit{J. Phys. A} \textbf{49} 184001 
	
	\bibitem{OV18} Oliver M and Vasylkevych S 2018 {Geodesic motion on the groups of diffeomorphisms with $H^1$ metric as geometric generalised Lagrangian mean theory} \textit{in preparation}  
	
	\bibitem{Palais} Palais R S 1968 \textit{Foundations of global nonlinear analysis} (New York: Benjamin) 
	
	\bibitem{RS06a} Roberts P H and Soward A M 2006 {Eulerian-Lagrangian means in rotating magnetohydrodynamic flows I. General results} \textit{Geophys. Astrophys. Fluid Dynam.} \textbf{100} 457--483
	
		\bibitem{RS06b} Roberts P H and Soward A M 2006 {Covariant description of non-relativistic magnetohydrodynamics} \textit{Geophys. Astrophys. Fluid Dynam.} \textbf{100} 485--502

\bibitem{Salmon13} Salmon R 2013 {An alternative view of generalized Lagrangian mean theory}. \textit{J. Fluid Mech.} \textbf{719} 165--182

\bibitem{Salmon16} Salmon R 2016 {Variational treatment of inertia-gravity waves interacting with a quasi-geostrophic mean flow} \textit{J. Fluid Mech.} \textbf{809} 502--529
	
	\bibitem{Shkoller00} Shkoller S 2000 {Analysis on groups of diffeomorphisms of manifolds with boundary and the averaged motion of a fluid} \textit{J. Diff. Geom.}  \textbf{55} 145--191.
	
	\bibitem{Shkoller02} Shkoller S 2002 {The Lagrangian averaged Euler (LAE-$\alpha$) equations with free-slip or mixed boundary conditions}, \textit{Geometry, Mechanics, and Dynamics} ed P Newton, P Holmes {\em et al}  (New York: Springer) 
	
\bibitem{SR10} Soward A M and Roberts P H 2010 {The hybrid Euler-Lagrange procedure using an extension of Moffatt's method} \textit{J. Fluid Mech.} \textbf{661} 45--72 

\bibitem{SR14} Soward A M and Roberts P H 2014 {Eulerian and Lagrangian means in rotating, magnetohydrodynamic flows II. Braginsky's nearly axisymmetric dynamo} \textit{Geophys. Astrophys. Fluid Dynam.} \textbf{108} 269--322

\bibitem{W15} Wagner G L and Young W R 2015 {Available potential vorticity and wave-averaged quasi-geostrophic flow} \textit{J. Fluid Mech.} \textbf{785} 401--424

\bibitem{X15} Xie J-H and Vanneste J 2015 {A generalised-Lagrangian-mean model of the interactions between near-inertial waves and mean flow} \textit{J. Fluid Mech.} \textbf{774} 143--169
\end{thebibliography}

 \end{document}